\documentclass[aps,prd,11pt,superscriptaddress,notitlepage,longbibliography,nofootinbib,tightenlines,groupedaddress]{revtex4-1}

\usepackage{amsmath,amssymb,amsfonts}
\usepackage{graphicx}
\usepackage{subfigure}
\usepackage{color}
\usepackage{hyperref}
\definecolor{darkred}{rgb}{0.8,0.1,0.1}
\hypersetup{colorlinks=true, linkcolor=darkred, citecolor=blue, linktoc=page}

\newcommand{\N}[1]{\ensuremath{\mathcal N=#1}}

\DeclareMathOperator{\tr}{tr}

\makeatletter\def\l@subsubsection#1#2{}%
\makeatletter\def\l@subsection#1#2{}%
\makeatother

\begin{document}

\title{Flavored \texorpdfstring{\N{4}}{N=4} SYM -- a highly entangled quantum liquid}

\author{Han-Chih Chang}
\email{hanchih@uw.edu}
\author{Andreas Karch}
\email{akarch@uw.edu}
\author{Christoph F.~Uhlemann}
\email{uhlemann@uw.edu}
\affiliation{Department of Physics, University of Washington, Seattle, WA 98195-1560, USA}

\begin{abstract}
We study \N{4} SYM theory coupled to fundamental \N{2} hypermultiplets in a state of finite charge density.
The setup can be described holographically as a configuration of D3 and D7 branes with a non-trivial
worldvolume gauge field on the D7.
The phase has been identified as a new form of quantum liquid,
where certain properties are those of a Fermi liquid while others are clearly distinct.
We focus on the entanglement among the flavors, as quantified by the entanglement entropy.
The expectation for a Fermi liquid would be a logarithmic enhancement of the area law, but
we find a more drastic enhancement instead.
The leading contributions are volume terms with a non-trivial shape dependence,
signaling extensive entanglement among the flavors.
At finite temperature these correlations are confined to a
region of size given by the inverse temperature.
\end{abstract}

\maketitle
\tableofcontents

\section{Introduction}
Exploring the possible quantum phases of matter has become a significant part of modern physics.
Established models like Fermi liquid theory do not account for the variety of phases
realized in condensed matter systems, and new concepts are called for.
Via the AdS/CFT dualities \cite{Maldacena:1997re,Witten:1998qj,Gubser:1998bc} string theory has become
a useful tool to map out the landscape of quantum phases of matter, and in this work we study one specific example.
We focus on \N{4} supersymmetric Yang-Mills theory (SYM) coupled to \N{2} flavor hypermultiplets in a state of finite charge density.
In the sense of being a translationally invariant zero-temperature state of finite charge density, this
is indeed a quantum liquid.
In fact, a rather interesting one \cite{Karch:2008fa}: despite a zero-sound mode with the properties
expected for a Fermi liquid, other properties like the specific heat clearly characterize it as a non-Fermi liquid.
It also exhibits a degenerate ground state, signaled by a non-vanishing entropy density at zero temperature,
which is a feature often associated with topological order \cite{Haldane:1985eda,Wen:1990zza},

The study of entanglement has emerged as a powerful diagnostic to detect and characterize different phases of matter,
which in particular includes non-Fermi liquid behavior  \cite{2014arXiv1403.0577S}.
Entanglement entropy as the most prominent quantification is used, e.g., to
study quantum critical phenomena \cite{Vidal:2002rm}
and to detect topological order \cite{Kitaev:2005dm,Levin:2006zz}.
The usual definition in quantum field theory rests on a geometric split
into two subsystems $A$ and $B$ at a constant time.
For a global state described by a density operator $\rho$, it is then defined as the von-Neumann entropy
of the reduced density operator $\rho_A$ for, say, the subsystem $A$\footnote{%
For gauge theories, the geometric split does not yield a tensor decomposition of the Hilbert space, but
an entanglement entropy can still be defined \cite{Casini:2013rba}.}
\begin{align}\label{eqn:SEE-definition}
 \mathcal S_\mathrm{EE}&=-\tr \rho_A\log\rho_A~,& \rho_A&=\tr_B\rho~.
\end{align}
We use it here to study correlations among the charge carriers in flavored \N{4} SYM to better understand the phase
of finite charge density.
Our results may also be of more general interest, as many of the investigations of entanglement entropy
have focused on the vacuum and analytical results for more general states are still rather scarce.

The actual evaluation of (\ref{eqn:SEE-definition}) for QFTs using the replica trick is non-trivial,
even for free theories.
Fortunately, we can exploit the fact that the computation is greatly simplified in the
dual gravity description as obtained from AdS/CFT.
In that context, the  \N{4} SYM coupled to fundamental \N{2} flavor hypermultiplets
corresponds to a configuration of intersecting D3/D7 branes \cite{Karch:2002sh}.
The finite charge density is realized by a non-trivial background profile for the D7 worldvolume gauge
field corresponding to the diagonal U($1$) of the flavor symmetry group \cite{Kobayashi:2006sb,Karch:2007pd,Karch:2007br,Mateos:2007vc}.
The quenched approximation on the field theory side corresponds to weakly backreacting D7 branes.
Computing entanglement entropies holographically simply amounts to evaluating the area of a minimal surface
in AdS \cite{Ryu:2006bv,Lewkowycz:2013nqa}, which is strikingly simpler than the corresponding field-theory calculation.
Nevertheless, in the present context even that becomes tough.
To obtain the flavor contribution to the entanglement entropy, one would first have to compute the backreaction
of the D7 branes with the worldvolume gauge field on the background geometry created by the D3 branes.
For D7 branes homogeneously smeared over the compact part of the background geometry
that backreaction has been obtained in \cite{Bigazzi:2011it,Bigazzi:2013jqa}.
In the perturbed geometry one would then have to find the appropriate minimal surface and compute its area.
For the case of massive flavors at zero density this could be done analytically in \cite{Kontoudi:2013rla}.

Recently, two holographic methods have been developed \cite{Chang:2013mca,Karch:2014ufa}, which 
avoid dealing with the backreaction explicitly and allow us to make progress more easily.
For the calculations here we start from the method of \cite{Chang:2013mca}.
It only needs the linearized backreaction of an effective source on the AdS$_5$ part of the
background geometry created by the D3 branes.
The change in the minimal area can then be obtained as an integral over the minimal surface in the unperturbed
geometry.
To obtain results in a decent closed form for the state of finite charge density, we further exploit that
representation. This eventually allows us to avoid an explicit computation of the backreaction altogether.

The paper is organized as follows.
We review the holographic setting in Section \ref{sec:D3/D7-gauge-field} and set up Einstein's equations for
the effective backreaction of the D7 branes on the non-compact part of the geometry.
In Section \ref{sec:EE-zero-T} we calculate the entanglement entropies for a disc and a strip in closed form
as simple one-dimensional integrals.
We discuss the limits of small and large entangling surfaces in Section \ref{sec:entanglement-thermodynamics}.
We recover the universal entanglement temperature predicted for small regions and argue that the specific
form of the volume terms found for large regions signals extensive entanglement.
In Section \ref{sec:finite-temperature} we consider finite temperature and discuss the transition to the thermal entropy.
A summary and discussion of our results is given in the final Section \ref{sec:discussion}.

\section{D3/D7 with worldvolume gauge field: effective Einstein equations}\label{sec:D3/D7-gauge-field}
We start with a review of the setup used in \cite{Kobayashi:2006sb,Karch:2007pd,Karch:2007br,Mateos:2007vc} to
holographically describe flavored \N{4} SYM at finite charge density.
The discussion of the D3/D7 system straightforwardly generalizes to probe branes with a $p\,{+}\,1$ dimensional 
worldvolume in AdS$_{d+1}$ times an internal space. 
For the D3/D7 system the corresponding AdS$_5$ $\times$ $S^5$ space is created by the D3 branes and the D7 branes are embedded as probes. 
More generally, we consider a black-brane background 
which is a product of an asymptotically-AdS$_{d+1}$ black hole with metric
\begin{align}\label{eqn:D3-finite-T}
 ds^2&=\frac{L^2}{z^2}\left[\frac{dz^2}{b(z)}-b(z)dt^2+d\vec{x}^2\right]~,&b(z)&=1-\frac{z^{d}}{z_h^{d}}~,
 &z_h&=\frac{d}{4\pi T}~,
\end{align}
and an internal space $X$.
In this background we consider probe D$p$-branes wrapping the entire AdS$_{d+1}$ part.
The Dirac-Born-Infeld (DBI) action governing the dynamics of the D$p$ probe involves a U($N_f$) gauge field, of which we only need the diagonal U($1$).
To study the dual theory at finite density, we add a non-trivial profile for $A_t(z)$.
The DBI action in mostly-plus signature then reads
\begin{align}\label{eqn:brane-action}
 S_\mathrm{brane}&=-N_fT_p\int d^{p^\prime+1}y \sqrt{-\det(\gamma+2\pi\alpha^\prime F)}~.
\end{align}
The relation of $N_f$ and $T_p$ to field-theory quantities is given, e.g., in \cite{Karch:2008fa}.
For massless flavors, the branes wrap a constant part of the internal space and we can just integrate
that out. This yields a factor of the volume in the internal space, $V_X$, and we arrive at
\begin{align}\label{eqn:brane-action-5d}
S_\mathrm{brane}&=-T_0\int d^{d+1}y_s \sqrt{-\det(\gamma_s+2\pi\alpha^\prime A_t^\prime dz\wedge dt)}~,
\end{align}
where $T_0=N_fT_p V_X$.
We drop the subscript $s$ in the following.
The action does not depend on $A_t$ itself, only on the radial derivative. Hence we get a conserved quantity $q$,
which corresponds to the charge density in the dual field theory.
We absorb a factor $2\pi\alpha^\prime$ and define it by
\begin{align}
 2\pi\alpha^\prime\,T_0 q&:=\frac{\delta S_\mathrm{brane}}{\delta A_t'}
 =T_0 \sqrt{-\gamma}\frac{(2\pi\alpha^\prime)^2 F^{tz}}{\sqrt{1+(2\pi\alpha^\prime)^2 F^{zt}F_{zt}}}~.
\end{align}
This can be solved straightforwardly for $A_t^\prime$.
We now turn to setting up Einstein's equations for the effective backreaction of (\ref{eqn:brane-action-5d})
on the AdS$_5$ part of the background geometry, (\ref{eqn:D3-finite-T}).
For more detailed discussions of the D3/D7 setup we refer to \cite{Kobayashi:2006sb,Karch:2007pd,Karch:2007br,Mateos:2007vc}.
The effective energy-momentum tensor for (\ref{eqn:brane-action-5d}) reads
\begin{align}\label{eqn:Tmunu-general}
 T^{\mu\nu}&=\frac{2}{\sqrt{-g}}\frac{\delta S_\mathrm{brane}}{\delta g_{\mu\nu}}
 =-T_0\sqrt{\det\left[g^{-1}(\gamma+2\pi\alpha^\prime F)\right]}\left(\gamma+2\pi\alpha^\prime F\right)^{-1\,\lbrace\mu\nu\rbrace}~.
\end{align}
Here we only consider the effective theory after integrating out the internal space. As we will review in the next section, this has been shown to be sufficient to capture the leading order correction to the entanglement entropy \cite{Chang:2013mca}.
The indices of $(\gamma+2\pi\alpha^\prime F)^{-1}$ are symmetrized since $\delta g_{\mu\nu}$ is symmetric by construction.
The only non-vanishing components are
\begin{align}\label{eqn:Tmunu-eff}
  T^{zz}&=-T_0 g^{zz}\sqrt{1+q^2z^{2d-2}}~,&
  T^{tt}&=-T_0 g^{tt}\sqrt{1+q^2z^{2d-2}}~,&
  T^{ij}&=-\frac{T_0 g^{ij}}{\sqrt{1+q^2z^{2d-2}}}~.
\end{align}
Anticipating that the linearly backreacted geometry will be asymptotically AdS$_{d+1}$ with the same
radius of curvature as in the zero-density zero-temperature case, we choose the ansatz
\begin{align}\label{eqn:deltag-ansatz1}
 g+\delta g=\Big(1+\frac{t_0}{d(d-1)}\Big)g_{zz}dz^2+\big(1+t_0h(z)\big)g_{tt}dt^2+\big(1+t_0j(z)\big)g_{ij}dx^idx^j~,
\end{align}
where $t_0=\kappa T_0$ and $\kappa=16\pi G$.
We have also set $L=1$.
The non-trivial components of Einstein's equations are the $zz$, $tt$ and $ii$ components.
They only involve derivatives of $j$ and $h$, and to discuss their solutions we define $\tilde j$ and $\tilde h$ by
\begin{align}\label{eqn:jhtilde-def}
 \tilde j(z)&=z^{1-d}j^\prime(z)~,& \tilde h(z)&=z^{1-d}h^\prime(z)~.
\end{align}
The equations then simplify quite a bit and read
\begin{subequations}\label{eqn:Einstein-finiteT}
\begin{align}
  (d-1)z^d\left(b(z)\tilde h(z)+(d-1)b(z) \tilde j(z)-\frac{1}{2}zb^\prime(z)\tilde j(z)\right)&=\sqrt{1+q^2 z^{2d-2}}-1~,\label{eqn:Einstein-finiteT-zz}\\
  (d-1)z^{d+1}\sqrt{b(z)}\left(\sqrt{b(z)}\tilde j(z)\right)^\prime&=1-\sqrt{1+q^2 z^{2d-2}}~,\label{eqn:Einstein-finiteT-tt}\\
  z^{d+1} \left(b(z)\tilde h^\prime(z)+\frac{3}{2}b^\prime(z)\tilde h(z)+(d-2) \left(b(z)\tilde j(z)\right)^\prime\right)&=1-\frac{1}{\sqrt{1+q^2 z^{2d-2}}}~.
\end{align}
\end{subequations}
We can verify that these equations can indeed be solved with our ansatz as follows.
The first equation can be solved for $\tilde h$ in terms of $\tilde j$.
Using the result in the third equation then reproduces the second one.
Thus, once (\ref{eqn:Einstein-finiteT-tt}) is solved for $\tilde j$, this can indeed be extended to a full solution.
We impose regularity at the horizon by demanding $b \tilde h$ and $b \tilde j$ to vanish as $z\rightarrow z_h$.
Using this in either (\ref{eqn:Einstein-finiteT-zz}) or (\ref{eqn:Einstein-finiteT-tt}) both yields
\begin{align}\label{eqn:finite-T-jt-zh}
 \tilde j(z_h)&=\frac{2}{d(d-1)}z_h^{-d}\left(\sqrt{1+q^2z_h^{2d-2}}-1\right)~.
\end{align}

We focus on $T=0$ in the next sections, and come back to $T>0$ in Sec.~\ref{sec:finite-temperature}.
This means $b(z)=1$ and  the equations simplify further.
The condition (\ref{eqn:finite-T-jt-zh}) reduces to the demand that $\tilde j(z)$ vanishes as $z\rightarrow\infty$.
Using that, we integrate both sides of (\ref{eqn:Einstein-finiteT-tt}) on $[z,\infty)$, and find
\begin{align}\label{eqn:jtilde-sol}
 \tilde j(z)&=-\frac{1}{d(d-1)z^d}+\frac{q}{(d-1)z} {}_2F_1\Big(-\frac{1}{2},\frac{1}{2d-2},\frac{2d-1}{2d-2},-\frac{1}{z^{2d-2}q^2}\Big)~.
\end{align}
We did not find a closed expression for $j$, which is obtained by integrating $\tilde j$ with (\ref{eqn:jhtilde-def}).
We will therefore use integration by parts later to express the area in terms of $\tilde j$.
To this end we need the near-boundary expansion of $j$.
This can be obtained from the expansion of $\tilde j$, for which we find
\begin{align}\label{eqn:alpha-d}
  \tilde j(z)&=\frac{\alpha(d)}{d}q^{d/(d-1)}+\mathcal O(z^{d-2})~,  &
  \alpha(d)&=\frac{\Gamma \big(\frac{1}{2}-\frac{1}{2d-2}\big) \Gamma \big(1+\frac{1}{2d-2}\big)}{\sqrt{\pi }}~.
\end{align}
Integrating this further yields the near-boundary expansion of $j$
\begin{align}\label{eqn:j-near-boundary}
 j(z)&=c_2+c_1z^d+\mathcal O(z^{2d-2})~, & c_1&=\frac{\alpha(d)}{d^2}q^{d/(d-1)}~,
\end{align}
where the subleading parts are fixed and the remaining freedom is in the choice of $c_2$.
Eq.~(\ref{eqn:Einstein-finiteT-zz}) only fixes the derivative of $h$ in terms of the derivative of $j$,
so there is an analogous free constant in $h$.
To asymptotically get a Poincar\'{e}-AdS metric, we fix $h(0)=c_2=1/[d(d-1)]$.
This ensures that the perturbed geometry describes the perturbed dual CFT in the vacuum state,
i.e.\ we do not want to source $T_{tt}$ and $T_{ii}$.
Using (\ref{eqn:Einstein-finiteT-zz}) to relate the $\mathcal O(z^d)$ parts of $h$ and $j$, we find
\begin{align}\label{eqn:h-near-boundary}
 h(z)&=c_2-(d-1)c_1z^d+\mathcal O(z^{2d-2})~, & c_2&=\frac{1}{d(d-1)}~.
\end{align}

For the discussion of entanglement thermodynamics below we will also need the energy density
$\langle T_{tt}\rangle^{}_\mathrm{CFT}$ for the CFT at finite charge density.
It can be derived straightforwardly from the perturbed bulk metric, using
the one-point function of the renormalized CFT energy-momentum tensor as computed, e.g., in \cite{deHaro:2000xn}.
It reads
\begin{align}\label{eqn:finite-T-Tii}
 \langle T_{\mu\nu}\rangle^{}_\mathrm{CFT}&=\frac{d {L^\prime}^{d-1}}{16\pi G} g_{\mu\nu}^{(d)}~,
\end{align}
where $L^\prime$ is the radius of curvature of the perturbed solution, $L^\prime=L(1+t_0c_2/2)+\mathcal O(t_0^2)$.
With (\ref{eqn:deltag-ansatz1})
and the near-boundary expansions (\ref{eqn:j-near-boundary}), (\ref{eqn:h-near-boundary}),
we thus find, to linear order in $t_0$,
\begin{align}\label{eqn:energy-momentum-one-point}
\langle T_{tt}\rangle^{}_\mathrm{CFT}&=d(d-1)T_0c_1~,
&\langle T_{ii}\rangle^{}_\mathrm{CFT}&=dT_0c_1~.
\end{align}
We note that, using $\langle T_{ii}\rangle^{}_\mathrm{CFT}=p$,  we reproduce the thermodynamic pressure
$p=-(\partial\Omega/\partial V_{d-1})_{\mu,T}$ obtained from the potential $\Omega$ given in \cite{Karch:2008fa},
validating our backreaction.

We close the section with a comment on the validity of the linearized approximation.
The source term on the right hand side of Einstein's equations (\ref{eqn:Einstein-finiteT})
grows unboundedly in the IR. At zero temperature we can therefore not trust the linearized approximation
for the IR limit $z\,{\rightarrow}\,\infty$.
This point has been discussed in detail in \cite{Hartnoll:2009ns,Bigazzi:2013jqa}.
For the discussion of the entanglement entropy below, we will be interested in regions of large but
finite extent, such that the corresponding minimal surfaces have a finite extension into the bulk.
We thus do not need the actual IR limit and the backreaction in the region probed by the minimal
surfaces is small as long as $t_0$ is sufficiently small or, equivalently, if for a given fixed small $t_0$ the minimal surface is not parametrically large in $1/t_0$.

\section{Entanglement entropy at zero temperature}\label{sec:EE-zero-T}

In the following we compute the entanglement entropy at zero temperature.
To give the results in a clear form, we expand it as
\begin{align}\label{eqn:SEE-split}
 \mathcal S_\mathrm{EE}&=\mathcal S_\mathrm{EE}^{(0)}+\mathcal S^{(1)}_\mathrm{EE}+\mathcal O(t_0^2)~,
 &
  \mathcal S^{(1)}_\mathrm{EE}&=\mathcal S^{(1)}_\mathrm{EE,\,q=0}+\Delta\mathcal S^{(1)}_\mathrm{EE}~.
\end{align}
That is,  $\mathcal S^{(i)}_\mathrm{EE}$ denotes the $\mathcal O(t_0^i)$ contribution, but we have not
extracted the power of $t_0$ explicitly.
Moreover, we have isolated the renormalized entanglement entropy $\Delta \mathcal S_\mathrm{EE}$,
obtained by subtracting off the entanglement entropy in the vacuum state.
The leading contribution to the renormalized entropy is $\Delta\mathcal S^{(1)}_\mathrm{EE}$.
Following \cite{Chang:2013mca}, the $\mathcal O(t_0)$ change in the entanglement entropy due to the backreaction
of the flavor branes can be expressed as an integral over the minimal surface in the unperturbed geometry
\begin{align}\label{eqn:SEE-eff}
 \mathcal S^{(1)}_\mathrm{EE}&=\frac{1}{4G}\int \frac{1}{2}T_\text{min}^{\mu\nu} \delta g_{\mu\nu}~,&
T_\mathrm{min}^{\mu\nu}&=\frac{2}{\sqrt{-g}}\frac{\delta A_\mathrm{min}^{(0)}}{\delta g_{\mu\nu}}~,
\end{align}
where $A_\mathrm{min}^{(0)}$ denotes the area of the minimal surface in the unperturbed geometry.
This would usually involve an $8$-dimensional minimal surface and the backreaction in the entire $10$-dimensional spacetime.
Nicely enough, though, as shown in \cite{Chang:2013mca} using a detour via a double-integral formula,
the details of the internal space can be subsumed into an effective source on the AdS part.
In our setting this is just (\ref{eqn:brane-action-5d}), and with the discussion of the effective backreaction in the previous
section, we can now evaluate (\ref{eqn:SEE-eff}) for a spherical region and a strip.

\subsection{Disc}
To compute the $\mathcal O(t_0)$ entanglement entropy for the spherical region $A:|\vec{x}|\leq\ell$, we switch to spherical coordinates such that
$g_{ij}dx^idx^j=z^{-2}(dr^2+r^2d\Omega_{d-2}^2)$\,.
The original minimal surface can then be parametrized by $z=\ell s$, $r=\ell\sqrt{1-s^2}$
and is the hyperbolic space $\mathbb{H}^{d-1}$.
This yields the induced metric on the minimal surface
\begin{align}\label{eqn:induced-metric-minimal-surface}
 \gamma^{}_\mathrm{min}&=\frac{L^2}{s^2}\left(\frac{ds\otimes ds}{1-s^2}+(1-s^2)\,g^{}_{\mathrm{S}^{d-2}}\right)~.
\end{align}
The variation of the minimal area with respect to the bulk metric for $L=1$ yields
\begin{align}\label{eqn:Tmin}
 T_\mathrm{min}&=
 s^2\ell^2\left(\sqrt{1-s^2}\partial_z-s\partial_r\right)\otimes\left(\sqrt{1-s^2}\partial_z-s\partial_r\right)
 +g_{\mathrm{S}^{d-2}}^{ij}\frac{s^2}{1-s^2}\partial_i\otimes\partial_j~.
\end{align}
With the ansatz (\ref{eqn:deltag-ansatz1}) for the perturbed bulk metric, the change in the minimal area, (\ref{eqn:SEE-eff}), becomes
\begin{align}\label{eqn:SEE1}
 \mathcal S^{(1)}_\mathrm{EE}&=
\frac{t_0 V_{\mathrm{S}^{d-2}}}{4G}\int_{\epsilon/\ell}^1ds \frac{(1-s^2)^\frac{d-3}{2}}{2s^{d-1}}\left((s^2+d-2)j(\ell s)+\frac{1-s^2}{d(d-1)}\right)~.
\end{align}
To compute the separate parts according to (\ref{eqn:SEE-split}), we note that $j(\ell s)\vert_{q=0}=c_2$.
This yields
\begin{align}
 \mathcal S_\mathrm{EE,q=0}^{(1)}&=\frac{t_0}{2d}\frac{V_{\mathrm{S}^{d-2}}}{4G}\int_{\epsilon/\ell}^1 ds\,s^{1-d}(1-s^2)^\frac{d-3}{2}
 =\frac{t_0}{2d}\frac{V_{\mathbb{H}^{d-1}}}{4G}~.
\end{align}
In the last equality we have introduced the regularized volume of the hyperbolic space which is the original minimal surface.
This nicely reproduces the zero-density results of \cite{Chang:2013mca,Jensen:2013lxa}.
The remaining part is the renormalized entanglement entropy, which becomes
\begin{align}
 \Delta\mathcal S^{(1)}_\mathrm{EE}&=
 \frac{t_0 V_{\text{S}^{d-2}}}{4G}\int_{\epsilon/\ell}^1ds \frac{(1-s^2)^\frac{d-3}{2}}{2s^{d-1}}(s^2+d-2)\left(j(\ell s)-c_2\right)~.
\end{align}
To evaluate it without having to solve for the backreaction explicitly, we use integration by parts, which yields
\begin{align}\label{eqn:deltaS-double-integral}
\frac{4G}{t_0 V_{\text{S}^{d-2}} }\Delta\mathcal S^{(1)}_\mathrm{EE}&=
  \frac{\ell^{d-2}}{2\epsilon^{d-2}}\left[1-\frac{\epsilon^2}{\ell^2}\right]^{\frac{d-1}{2}}\left(j(\epsilon)-c_2\right)
  +\frac{\ell^d}{2}\int_{\epsilon/\ell}^1ds\,s  (1-s^2)^\frac{d-1}{2}\tilde j(\ell s)~.
\end{align}
In the second term we have expressed $j^\prime(\ell s)$ in terms of $\tilde j(\ell s)$ using (\ref{eqn:jhtilde-def}).
From the near-boundary expansion (\ref{eqn:j-near-boundary}), we see that $j(\epsilon)-c_2=\mathcal O(\epsilon^d)$,
such that the first term vanishes and we are only left with the second one.
The integrand is finite for $s\rightarrow 0$, such that we can equivalently set the lower bound of integration to zero.
With $\tilde j$ given in (\ref{eqn:jtilde-sol}) our final result then becomes
\begin{align}\label{eqn:deltaS-double-integral-2}
\Delta\mathcal S^{(1)}_\mathrm{EE}&=\frac{t_0 }{8G}V_{\text{S}^{d-2}}
  \int_{0}^1ds\,s  (1-s^2)^\frac{d-1}{2}\ell^d\tilde j(\ell s)~.
\end{align}
We note that $\ell^d\tilde j(\ell s)$ and thus $\Delta\mathcal S^{(1)}_\mathrm{EE}$ depend on $q$ and $\ell$ only through the
dimensionless combination $q\ell^{d-1}$.

\subsection{Strip}
We now turn to the entanglement entropy for a strip defined by $|x^1|\,{\leq}\,\ell/2$ at $t\,{=}\,0$.
Using Cartesian coordinates, $g_{ij}dx^i dx^j=z^{-2}d\vec{x}^{\,2}$,
the corresponding minimal surface can
be parametrized by $z$, $x^1\,{=}\,x^1(z)$ and $x^{i_0}$, ${i_0}\,{=}\,2,..,d\,{-}\,1$.
As discussed already in \cite{Ryu:2006bv}, extremizing the area in AdS$_{d+1}$ leads to
\begin{align}\label{eqn:slab-minimal-surface}
 \frac{dx^1}{dz}&=\pm \frac{1}{\sqrt{(\ell_\star/z)^{2d-2}-1}}~,&
 \ell_\star&=\frac{\ell}{2\sqrt{\pi}}\frac{\Gamma(\frac{1}{2d-2})}{\Gamma(\frac{d}{2d-2})}~,
\end{align}
where $\ell_\star$ marks how far the minimal surface extends into the bulk.
The induced metric then is
\begin{align}\label{eqn:slab-gamma-min}
  \gamma_\mathrm{min}&=\frac{L^2}{z^2}\left(\frac{dz\otimes dz}{1-(z/\ell_\star)^{2d-2}}+\sum_{i=2}^{d-1}dx^i\otimes dx^i\right)~.
\end{align}
Since $\delta g$ is diagonal, we only need the diagonal elements of $T_\mathrm{min}$, and they evaluate to
\begin{align}\label{eqn:Tmin-slab}
T_\mathrm{min}^{zz}&=\gamma_\mathrm{min}^{zz}~,&
T_\mathrm{min}^{11}&=\frac{\ell_\star^2}{L^2}\frac{z^{2d}}{\ell_\star^{2d}}~,&
T_\mathrm{min}^{i_0j_0}&=\gamma_\mathrm{min}^{i_0j_0}~.&
\end{align}
The $\mathcal O(t_0)$ entanglement entropy correction (\ref{eqn:SEE-eff}) with $L=1$ then becomes
\begin{align}\label{eqn:EE-slab-1}
 \mathcal S^{(1)}_\mathrm{EE}&=
 \frac{t_0V_{d-2}}{4G}\int_\epsilon^{\ell_\star} dz\sqrt{\gamma_\mathrm{min}}\left[
 \frac{1-(z/\ell_\star)^{2d-2}}{d(d-1)}+j(z)\left(\frac{z^{2d-2}}{\ell_\star^{2d-2}}+d-2\right)
 \right]~.
\end{align}
The $(d\,{-}\,2)$-dimensional volume of the strip in the transverse directions is denoted by $V_{d-2}$,
and the integration over $z$ from $\epsilon$ to $\ell_\star$ and back gives a factor $2$.
We now perform a change of variables to $v=z/\ell_\star$ and slightly rearrange the terms in square brackets,
to arrive at
\begin{align}\label{eqn:SEE-slab-2}
 \mathcal S^{(1)}_\mathrm{EE}&=
 \frac{t_0V_{d-2}}{4G}\ell_\star^{2-d}\int_{\epsilon/\ell_\star}^{1} dv\frac{v^{1-d}}{\sqrt{1-v^{2d-2}}}\left[
 \frac{1}{d}+(j(v\ell_\star)-c_2)\left(v^{2d-2}+d-2\right)
 \right]~.
\end{align}
The first term in square brackets yields the zero-density result, and the second one the renormalized entropy.
As a consistency check and for later reference we also evaluate the former, which yields
\begin{align}\label{eqn:SEE-strip-q0}
 \mathcal S^{(1)}_\mathrm{EE,\,q=0}&=
 \frac{t_0}{2d}\frac{1}{4G}\frac{V_{d-2}}{d-2}\left[ \frac{2}{\epsilon^{d-2}}
  -\ell^{2-d}\left(2\sqrt{\pi}\frac{\Gamma(\frac{d}{2(d-1)})}{\Gamma(\frac{1}{2(d-1)})}\right)^{d-1}
 \right]~.
\end{align}
This is the result found in \cite{Ryu:2006bv} up to a factor $t_0/(2d)$, in accordance with \cite{Chang:2013mca,Jensen:2013lxa}.
We can now turn to the finite-density contribution, given by the second term in the square brackets in (\ref{eqn:SEE-slab-2}).
Using integration by parts again, we obtain
\begin{align}\label{eqn:Delta-strip-1}
  \frac{4G}{t_0V_{d-2}}\Delta\mathcal S^{(1)}_\mathrm{EE}&=-\left[\left(j(v\ell_\star)-c_2\right)v^{2-d}\sqrt{1-v^{2d-2}}\right]_{\epsilon/\ell_\star}^1
         +\ell_\star^{2}\int_{\epsilon/\ell_\star}^1 dv\, \tilde j(v \ell_\star)v\sqrt{1-v^{2d-2}}~.
\end{align}
Since $j(z)-c_2=\mathcal O(z^d)$, the first term does not contribute as $\epsilon\,{\rightarrow}\,0$,
and we are left with the second one.
The lower bound of integration again can be set to zero since the integrand is finite.
With $\tilde j$ given in (\ref{eqn:jtilde-sol}) and $\ell_\star$ in (\ref{eqn:slab-minimal-surface}) the final result reads
\begin{align}\label{eqn:Delta-strip-2}
 \Delta\mathcal S^{(1)}_\mathrm{EE}&=\frac{t_0}{4G}V_{d-2}\ell_\star^{2}\int_{0}^1 dv\, \tilde j(v \ell_\star)v\sqrt{1-v^{2d-2}}~.
\end{align}
Due to the factor $V_{d-2}$, the entanglement entropy for the strip is not a function of $q\ell^{d-1}$ alone,
but we note that the rescaled entropy $\ell^{d-2}\Delta\mathcal S^{(1)}_\mathrm{EE}$ is.

\section{Entanglement thermodynamics and shape-dependent volume terms}\label{sec:entanglement-thermodynamics}
In this section we first derive the behavior of the entanglement entropy for the limits of small and large entangling surfaces,
which means the narrow and wide limits for the strip.
We then discuss the behavior for small regions in the context of the
$1^\mathrm{st}$ law of entanglement thermodynamics, and relate the result for large
regions to the entropy of the global state.
The expressions for small regions can be obtained straightforwardly by expanding (\ref{eqn:deltaS-double-integral-2})
and (\ref{eqn:Delta-strip-2}) for small $q \ell^{d-1}$.
The expansion of $\tilde j$ for small argument is given in (\ref{eqn:alpha-d}), and the entanglement entropies
(\ref{eqn:deltaS-double-integral-2}) and (\ref{eqn:Delta-strip-2}) then evaluate to
\begin{align}\label{eqn:SEE-small}
 \Delta \mathcal S_\mathrm{EE}^{(1)}&=\frac{t_0}{2d}\frac{V_{S^{d-2}}}{4G}\frac{\alpha(d)}{d+1}q^\frac{d}{d-1}\ell^d~,
 &
 \Delta \mathcal S_\mathrm{EE}^{(1)}&=\frac{t_0}{2d}\frac{V_{d-2}}{4G}\ell_\star^2q^\frac{d}{d-1}  \frac{\sqrt{\pi}\alpha(d)\Gamma(\frac{1}{d-1})}{(d+1)\Gamma(\frac{1}{2}+\frac{1}{d-1})}~,
\end{align}
for the sphere and the strip, respectively.

For large entangling surface the analogous expansion is less straightforward,
and we work it out in some detail for the sphere.
The point is that, no matter how large $q\ell^{d-1}$ is,
$q(s\ell)^{d-1}$ is always small in some part of the integration region,
so we can not simply expand the integrand in (\ref{eqn:deltaS-double-integral-2}).
To still obtain a systematic expansion we first perform one more integration
by parts in (\ref{eqn:deltaS-double-integral-2}), such that the resulting integral only involves $\tilde j^\prime$.
Using $\tilde j(0)=d c_1$ and (\ref{eqn:Einstein-finiteT-tt}) to eliminate $\tilde j^\prime$,
this yields
\begin{align}\label{eqn:deltaS-double-integral4}
\Delta\mathcal S_\mathrm{EE}^{(1)}&=\frac{t_0V_{\text{S}^{d-2}} }{8(d+1)G}\left[\ell^{d}dc_1
+\int_{0}^1 ds \frac{(1-s^2)^\frac{d+1}{2}}{(d-1)s^{d+1}}\left(1-\sqrt{1+q^2 (\ell s)^{2d-2}}\right)\right]~.
\end{align}
We now split the integral using $[0,1]=[0,\eta)\cup [\eta,1]$ with $\eta\ll 1$.
On $[\eta,1]$, $s$ is bounded from below by $\eta$, so we can expand the square root in (\ref{eqn:deltaS-double-integral4}).
On $[0,\eta]$  $s$ is small itself, so we can expand $(1-s^2)^{(d+1)/2}$.
The resulting integrals in both regions can then be done analytically.
Combining the results and expanding for small $\eta$ yields the desired expansion
\begin{align}\label{eqn:SEE-sphere-large-q}
 \Delta\mathcal S^{(1)}_\mathrm{EE}&=\frac{t_0}{4G}\left[\frac{q}{4d}\ell^{d-1}V_{\text{S}^{d-1}}
 -\frac{B(\frac{2d-3}{2d-2},\frac{d}{2d-2})}{8(d-2)} q^{\frac{d-2}{d-1}}\ell^{d-2}V_{\text{S}^{d-2}}+o(q^{\frac{d-2}{d-1}}\ell^{d-2})\right]~.
\end{align}
The volume of the $n$-dimensional unit sphere is $V_{\mathrm{S}^n}=2\pi^{(n+1)/2}/\Gamma(\frac{n+1}{2})$
and $B(a,b)$ is the Euler beta function.
The first term is a volume term on the boundary, and could also be obtained by na\"ively expanding the integrand
in (\ref{eqn:deltaS-double-integral-2}).
But the na\"ive expansion then breaks down and gives divergent subleading terms.
The noteworthy feature of the second term in (\ref{eqn:SEE-sphere-large-q}) is that it scales as just an area,
with no logarithmic enhancement.
For the strip we can analogously expand the entanglement entropy for large $q$.
The interesting part for the discussion below is the leading volume term, which reads
\begin{align}\label{eqn:SEE-strip-large-q}
 \Delta\mathcal S^{(1)}_\mathrm{EE}&=\frac{t_0}{4G}\frac{\ell_\star^2}{\ell^2}\frac{\pi}{d(d-1)}V_{d-2}\ell q+o(q\ell^{d-1})~.
\end{align}
We see that for the disc and the strip the entanglement entropies for large $\ell$ are proportional to
$q$ and the volume of the region $A$, $V_A$.
For the disc we have $V_A=\ell^{d-1}V_{\text{B}^{d-1}}$, with the volume of the unit ball
$V_{\text{B}^{n}}=\pi^{n/2}/\Gamma(1+\frac{n}{2})$, and for the strip $V_A=\ell V_{d-2}$.
Using this to work out the coefficients, they turn out to be different.
We will see this very explicitly in the entanglement thermodynamics discussion and come back
to the interpretation afterwards.

\begin{figure}
 \includegraphics[width=0.45\linewidth]{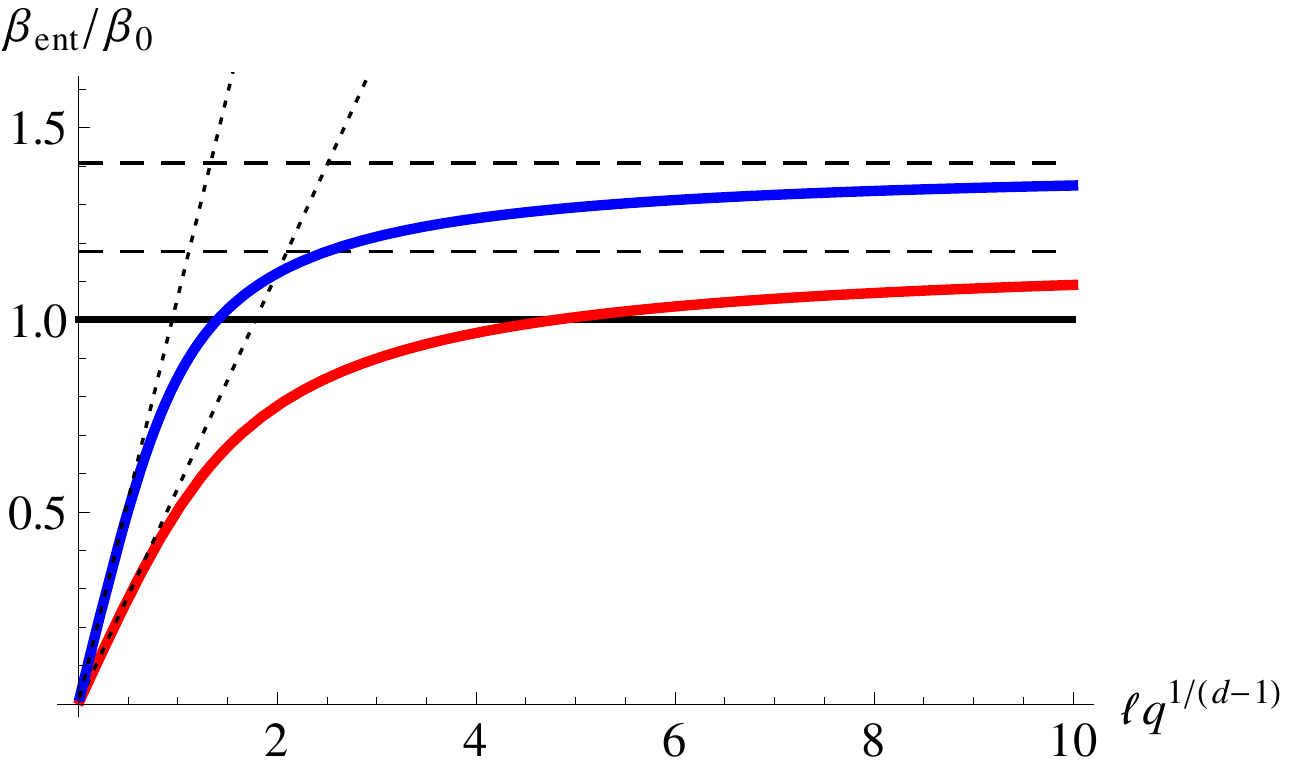}
 \caption{The upper blue and lower red curves show $\beta_\mathrm{ent}=1/T_\mathrm{ent}$
 with the entanglement temperature defined in (\ref{eqn:entanglement-temp}) for the strip and the disc, respectively, in $d\,{=}\,4$.
 Both are normalized to $\beta_0=\delta s_0/\delta E$, which is the value we would expect for large
 regions if the entanglement entropy were dominated by the entropy of the global state.
 The ratio only depends on $q\ell^{d-1}$ for the sphere and the strip, and $\beta_0$ is
 indicated as solid black line.
 The dotted lines show the shape-dependent but otherwise universal behavior expected for small regions,
 which the curves reproduce nicely.
 For larger regions we see a smooth cross over and the curves saturate at the constant values (\ref{eqn:beta-large-l})
 shown as dashed lines.
 The asymptotic values are different and greater than that expected due to the entropy of the global state in both cases,
 and the same applies for the entanglement entropy densities.\label{fig:EETemp}}
\end{figure}

\paragraph{Entanglement thermodynamics and universal behavior for small $\ell$}
We now discuss our results in the context of the first law of entanglement thermodynamics
proposed in \cite{Bhattacharya:2012mi}.
A field-theory derivation has been given in \cite{Wong:2013gua}.
This concept allows to define a notion of temperature also in situations where the usual thermodynamic one
is ill defined or trivial. We exploit this for our zero-temperature study.
The entanglement temperature $T_\mathrm{ent}$ is defined by
\begin{align}\label{eqn:entanglement-temp}
 T_\mathrm{ent}\delta \mathcal S_\mathrm{EE}(A)&=\delta E(A)~,&
 E(A)&=\int_A \langle T_{tt}\rangle^{}_\mathrm{CFT}~,
\end{align}
and we also define $\beta_\mathrm{ent}=1/T_\mathrm{ent}$.
To evaluate that definition we use (\ref{eqn:energy-momentum-one-point})
and (\ref{eqn:j-near-boundary}), which yields
\begin{align}\label{eqn:deltaE}
 \delta E(A) =T_0\alpha(d)V_A\,q^\frac{1}{d-1}\delta q~,
\end{align}
where $V_A=\ell^{d-1}V_{\text{B}^{d-1}}$ for the sphere and $V_A=\ell V_{d-2}$ for the strip.
For $q\ell^{d-1}\ll 1$ we then find from (\ref{eqn:SEE-small})
\begin{align}
 \beta^{\ell\ll 1}_\mathrm{ent}&=\frac{2\pi\ell}{d+1}~,&
 \beta^{\ell\ll1}_\mathrm{ent}&=\frac{\pi^{3/2} \Gamma(2+\frac{1}{d-1})}
 {d\,\Gamma (\frac{3}{2}+\frac{1}{d-1})}\frac{\ell_\star^2}{\ell}~,
\end{align}
for the sphere and the strip, respectively.
This is exactly the shape-dependent but otherwise universal behavior predicted for small regions in \cite{Bhattacharya:2012mi}.
It was interpreted as the fact that localized excitations
of the vacuum state store a universal amount of information, regardless of their precise nature.
We now turn to large regions, $q\ell^{d-1}\gg 1$.
Due to the volume terms in (\ref{eqn:SEE-sphere-large-q}), (\ref{eqn:SEE-strip-large-q})
and the fact that the energy density in (\ref{eqn:entanglement-temp}) is itself proportional
to the volume, the entanglement temperature settles on non-zero constant values for large $\ell$.
Concretely, we find for the sphere and the strip
\begin{align}\label{eqn:beta-large-l}
 \beta_\mathrm{ent}^{\ell\gg1}&=\frac{2\pi^{3/2}}{dq^{1/(d-1)}}\frac{\Gamma(\frac{d+1}{2})}{\alpha(d)\Gamma(\frac{d}{2})}~,&
 \beta_\mathrm{ent}^{\ell\gg1}&=\frac{2\pi^{3/2}}{dq^{1/(d-1)}}\frac{\Gamma(\frac{1}{2 d-2})}{\Gamma (\frac{d-2}{2d-2}) \Gamma (\frac{d}{2 d-2})^2}~,
\end{align}
respectively.
For intermediate sizes we can obtain $\beta_\mathrm{ent}$ numerically using (\ref{eqn:entanglement-temp}) and (\ref{eqn:deltaE})
with (\ref{eqn:deltaS-double-integral-2}) for the sphere and (\ref{eqn:Delta-strip-2}) for the strip.
The result is illustrated in Fig.~\ref{fig:EETemp}.
Despite the complicated appearance of the entanglement entropies, the results just smoothly interpolate
between our analytical results for small and large $\ell$.
In passing we also note that $\beta_\mathrm{ent}$ and thus also the renormalized entanglement entropies
increase monotonically with $\ell$ for fixed $q$.
This is consistent with the entanglement F-theorem for spherical regions in $d\,{=}\,3$ \cite{Casini:2012ei}.

\paragraph{Shape-dependent volume terms and the entropy of the global state}
For small regions we have seen that our results reproduce the universal
behavior predicted from entanglement thermodynamics, and we now come back to a
more detailed discussion of large regions and the volume terms in (\ref{eqn:SEE-sphere-large-q}), (\ref{eqn:SEE-strip-large-q}).
In the vacuum state there are no extensive terms in the entanglement entropy.
The correlations are local and the entropy of the reduced state $\rho_A$ is mainly due to ignored
correlations across the boundary of $A$. These are UV sensitive and the entropy thus divergent, which
yields the usual area law \cite{Srednicki:1993im}.
The finite parts for the disc and the strip are then at most $\mathcal O(\ell^0)$.
This changes when we consider a global state with non-vanishing entropy density, such as a thermal state.
For large regions the finite part of the entanglement entropy is then dominated by the entropy of
the global state. This contribution has a trivial shape dependence: for a translationally invariant system
it is simply the entropy density of the global state multiplied by the volume of $A$.
For our state of finite charge density the situations is not that different, as it also has
a non-vanishing entropy density $s_0=4\pi T_0 q/d$, even at zero temperature \cite{Karch:2008fa}.
We therefore should expect an extensive contribution in the finite parts, and it should have the trivial
shape dependence $\Delta\mathcal S_\mathrm{EE}=s_0 V_A+o(q\ell^{d-1})$.
Interestingly enough, though, this is not what we find.
The results of (\ref{eqn:SEE-sphere-large-q}), (\ref{eqn:SEE-strip-large-q})
show that the volume terms have a non-trivial shape dependence: they can not be understood
as a universal factor multiplying the volume of $A$.
This can also be seen in Fig.~\ref{fig:EETemp}. If the volume terms had a trivial shape dependence,
the inverse entanglement temperature for large regions would approach the universal value $\beta_0=\delta s_0/\delta E$
regardless of the shape, which is clearly not the case.
We also see that the inverse entanglement temperatures are both greater than $\beta_0$, so we are led to
conclude that there is a shape-dependent contribution in addition to the expected trivial part.
This shape dependence can hardly be due to the entropy of the global state, which
is translationally invariant and has no information on possible entangling surfaces.
It therefore signals the presence of long-range correlations, which are not confined to the entangling surface
but rather correlate the entire interior of the region $A$ with its complement.

\section{Transition to thermal entropy at finite temperature}\label{sec:finite-temperature}
We now turn to the case of finite temperature and study the fate of the extensive entanglement
correlations found in the previous section.
If the shape-dependent volume terms indeed signal long-range correlations, we would expect that
for an intermediate range of distances our previous results
remain valid, but over long enough distances thermal fluctuations should wash them out eventually.
For the renormalized entanglement entropy this means that, as we increase the size of the region $A$,
the coefficient of the volume term should at some point decrease to the thermal entropy density. 
This does not imply that the entropy itself has to be non-monotonic.
Indeed, we find a smooth cross over from an ``entanglement-dominated phase'' to a
``thermal phase'', during which the entropy still grows monotonically with the size of $A$.
We will focus on the strip, since we can then find the corresponding minimal surface at finite temperature
in a decent form.
We did not succeed in solving the Einstein equations analytically, but the problem can be formulated in a
form which makes it easily accessible for numerical methods.
We first derive the expression for the change in minimal area due to the backreaction
and then discuss how to numerically solve (\ref{eqn:Einstein-finiteT}).

As for $T\,{=}\,0$, we parametrize the codimension-$2$ minimal surface for the strip by $x^1=x^1(z)$,
and the induced metric then reads
\begin{align}\label{eqn:strip-gamma-min-finite-T}
 \gamma_\mathrm{min}&=\left(g_{zz}+g_{xx}x^\prime(z)^2\right)dz\otimes dz+g_{i_0j_0}dx^{i_0}\otimes dx^{j_0}~.
\end{align}
The integrand for the minimal area, $\sqrt{\gamma_\mathrm{min}}$, does not depend on
$x^1(z)$ and we get a conserved quantity $\ell_\star$, parametrizing the extension of
the minimal surface into the bulk.
This yields
\begin{align}\label{eqn:finite-T-slab}
 \frac{dx^1}{dz}&=\frac{z\sqrt{g_{zz}}}{\sqrt{(\ell_\star/z)^{2d-2}-1}}~.
\end{align}
The width $\ell$ of the strip is then obtained from its extension in the $x^1$-direction as
\begin{align}
 \ell=2x(0)&=\int^{\ell_\star}_0 \frac{2dz}{\sqrt{b(z)\left((\ell_\star/z)^{2d-2}-1\right)}}~.
\end{align}
We again only need the diagonal elements of $T_\mathrm{min}^{\mu\nu}$
and they take the form given in (\ref{eqn:Tmin-slab}), now with $\gamma_\mathrm{min}$ given in (\ref{eqn:strip-gamma-min-finite-T}).
The entanglement entropy correction (\ref{eqn:SEE-eff}) with (\ref{eqn:deltag-ansatz1}) then becomes
\begin{align}
 \mathcal S^{(1)}_\mathrm{EE}&=
 \frac{t_0V_{d-2}}{4G}\int_\epsilon^{\ell_\star} dz\sqrt{\gamma_\mathrm{min}}
 \left[\frac{g_{zz}T^{zz}}{d(d-1)}+j(z)\left(\frac{z^{2d-2}}{\ell_\star^{2d-2}}+d-2\right)\right]~.
\end{align}
Somewhat remarkably, the only difference to (\ref{eqn:EE-slab-1}) is in the volume form -- the redshift factor drops
out in the first term in square brackets and the second one does not include it either.
We substitute $v=z/\ell_\star$ and rewrite the expression as
\begin{align}\label{eqn:SEE-slab-finiteT}
 \mathcal S^{(1)}_\mathrm{EE}&=
 \frac{t_0V_{d-2}}{4G}\ell_\star^{2-d}\int_{\epsilon/\ell_\star}^{1} dv\frac{v^{1-d}}{\sqrt{b(v\ell_\star)(1-v^{2d-2})}}
 \left[
 \frac{1}{d}+(j(v\ell_\star)-c_2)\left(v^{2d-2}+d-2\right)
 \right]~.
\end{align}
The first term in the square brackets yields the zero-density result.
We define a finite quantity by subtracting off the entanglement entropy in the vacuum state
at zero charge and zero temperature
\begin{align}\label{eqn:stripEEren-finite-T}
 \Delta \mathcal S^{(1)}_\mathrm{EE}&=\mathcal S^{(1)}_\mathrm{EE}-\mathcal S^{(1)}_\mathrm{EE,q=T=0}~.
\end{align}
This is the expression we will evaluate numerically below.
Before turning to the results, we discuss the limit of an asymptotically wide strip.
In that case $\ell_\star\rightarrow z_h$ and we can get a simple expression for the entropy density.
The bulk minimal surface then approximately consists of two pieces:
one wraps the part of the horizon corresponding to the strip,
and the other connects that piece to the boundary.
The latter just contributes a constant, and only the former is therefore relevant
for the entropy density.
The induced metric is simply $\gamma_\mathrm{min}=g_{ij}dx^i\otimes dx^j$
and we find
\begin{align}\label{eqn:large-stripEE-finite-T}
 S^{}_\mathrm{EE,\infty}&=\frac{1}{4G}\int_{z=z_h}\sqrt{\gamma}=\frac{V_{d-1}}{4G}z_h^{1-d}\left(1+t_0 j(z_h)\right)^{(d-1)/2}
  +\mathcal O(V_{d-1}^0)+\mathcal O(t_0^2) ~.
\end{align}

\paragraph{Numerical solution and results}
We now turn to the evaluation of the renormalized entropy (\ref{eqn:stripEEren-finite-T}).
To this end we have to solve (\ref{eqn:Einstein-finiteT}) for the backreaction, which
can be done as follows.
With the initial value (\ref{eqn:finite-T-jt-zh}) we can integrate (\ref{eqn:Einstein-finiteT-tt}) to get $\tilde j$,
and then solve (\ref{eqn:Einstein-finiteT-zz}) for $\tilde h$.
To further integrate these using (\ref{eqn:jhtilde-def}), we have to fix two more constants of
integration. Since we do not want to source the CFT energy-momentum tensor, we once again fix $h(0)=j(0)=c_2$.
We note that the temperature of the backreacted solution is not necessarily the same as that
of (\ref{eqn:D3-finite-T}). This should not concern us at the moment, but will be relevant
for the cross check with thermodynamic results below.

\begin{figure}[ht]
\center
\subfigure[][]{
  \includegraphics[height=0.30\linewidth]{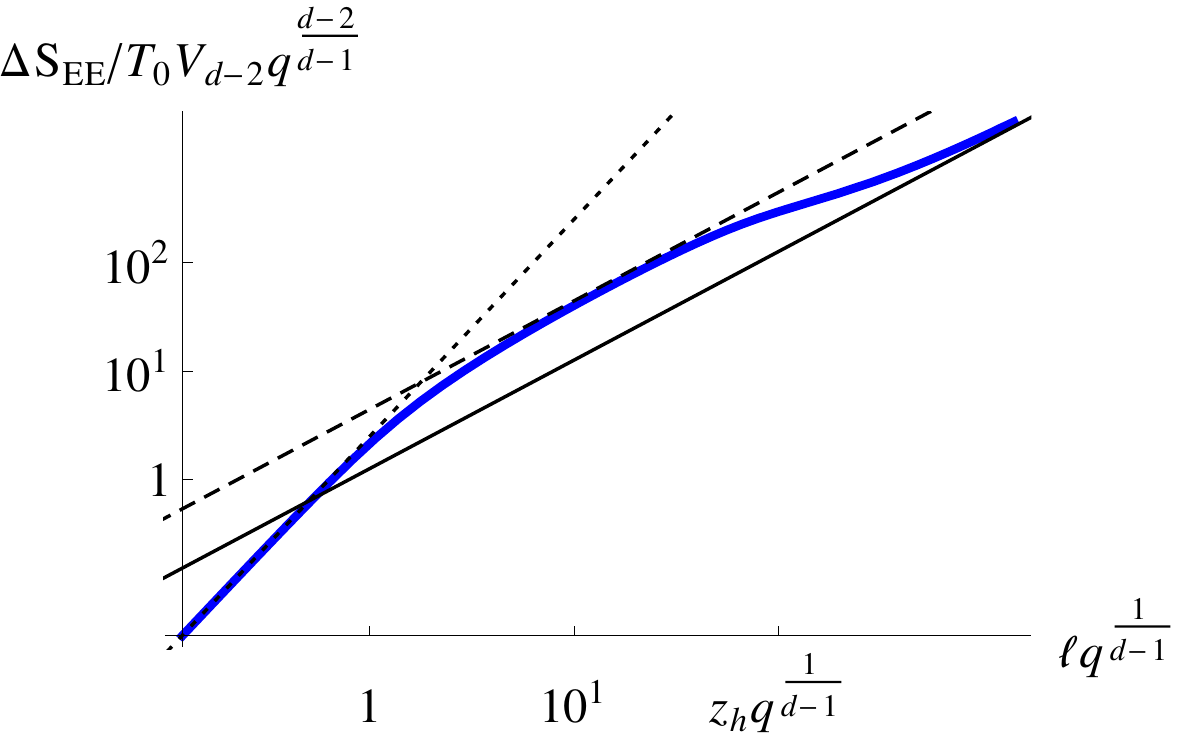} \label{fig:EntropyTransition}
}\qquad
\subfigure[][]{ \label{fig:entropy}
    \includegraphics[height=0.29\linewidth]{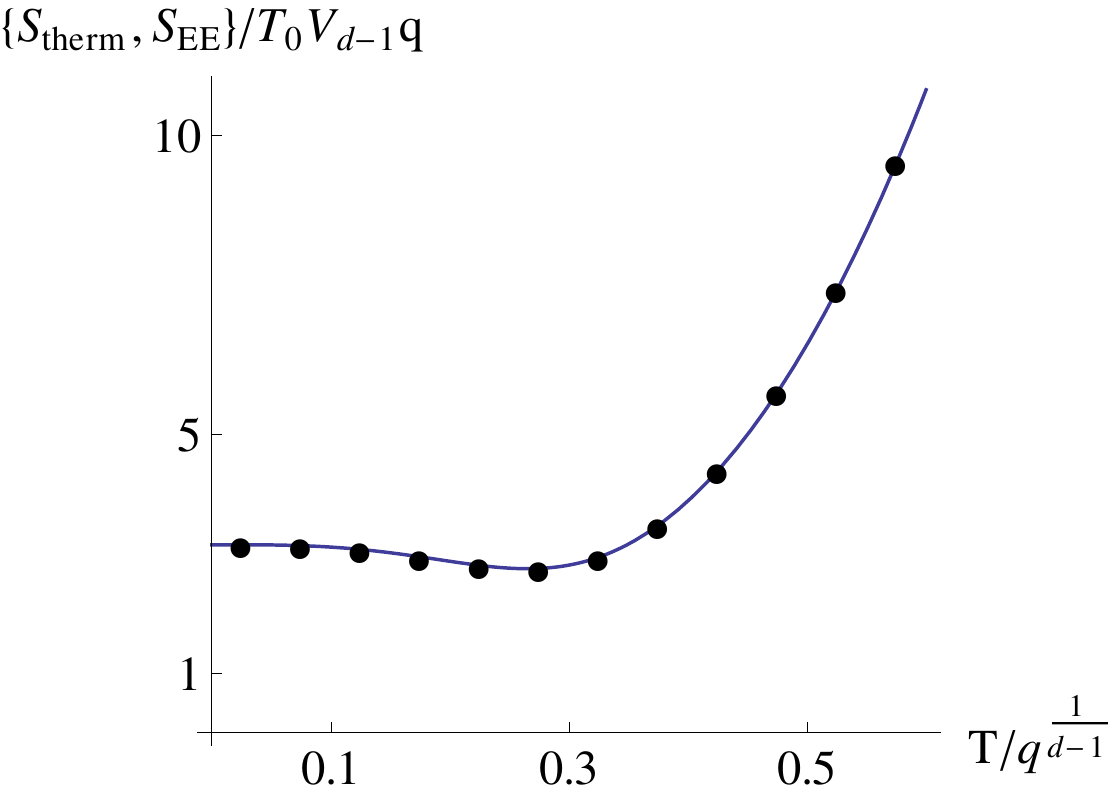}
}
\caption{
The figure on the left hand side shows as blue solid curve the entanglement entropy of the strip,
(\ref{eqn:stripEEren-finite-T}).
The rescaling makes it a function of only the dimensionless quantities $q\ell^{d-1}$
and $T/q^{\frac{1}{d-1}}$.
We have chosen $d\,{=}\,4$ and a small temperature such that $z_hq^{\frac{1}{d-1}}=10^{2}$.
The dotted and dashed lines directly correspond to those of Fig.~\ref{fig:EETemp}:
they show the zero-temperature results for the narrow and the wide strip, respectively.
The black solid line shows the entropy for the infinitely extended strip, (\ref{eqn:large-stripEE-finite-T}),
which is expected to reproduce the entropy of the global state.
That this is indeed the case is shown on the right hand side:
the solid line shows the entropy of the global state,
and the markers show the entanglement entropy for the asymptotically large strip
as given in (\ref{eqn:EE-strip-large-old-T}).
We conclude that the entanglement entropy is well described by the zero-temperature 
result as long as the minimal surface does not extend all the way to the horizon.
For even wider strips, the shape-dependent enhancement vanishes, leaving only the entropy of the global state.\label{fig:finiteT}
}
\end{figure}

With the backreaction at hand, we can then straightforwardly evaluate (\ref{eqn:stripEEren-finite-T})
with (\ref{eqn:SEE-slab-finiteT}).
The results are illustrated in Fig.~\ref{fig:EntropyTransition}, where we have chosen $d=4$ as
appropriate for \N{4} SYM.
For small $\ell$ we see that the entanglement entropy agrees very well with the zero-temperature analysis.
The figure also confirms our expectation for large $\ell$:
For $\ell\gg1$ but still small enough such that $\ell_\star$ is sufficiently far from the horizon,
we have the shape-dependent contribution to the volume law.
At around the point where $\ell$ equals $z_h$, the minimal surface starts to really probe the
horizon (note the linear relation between $\ell$ and $\ell_\star$ for $T=0$).
As $\ell$ is increased further, the renormalized entropy finally settles on the universal volume
contribution due to the entropy of the global state.
This shows that the thermal fluctuations wash out the long-range correlations on scales greater
than $z_h$.


As a cross check for our numerical implementation we explicitly verify that the asymptotic form of the entanglement entropy for the strip,
(\ref{eqn:large-stripEE-finite-T}), reproduces the entropy density of the global state.
To this end we first have to determine the temperature of the perturbed solution,
as usually by demanding that there should be no conical singularity in the Euclidean solution.
To linear order in $t_0$ this yields
\begin{align}\label{eqn:deltaT}
 T&=\frac{d}{4\pi z_h}\left[1+t_0\left(\frac{h(z_h)-c_2}{2}-\frac{\delta z_h}{z_h}\right)\right]+\mathcal O(t_0^2)~,
\end{align}
where we have defined the locus of the horizon in the perturbed solution as $z_h^\prime=z_h+t_0\delta z_h+\mathcal O(t_0^2)$.
To keep the temperature unchanged we thus have to fix $\delta z_h=z_h(h(z_h)-c_2)/2$.

We can now compare the asymptotic behavior of the entanglement entropy
to the corresponding entropy density of the global state.
We expand (\ref{eqn:large-stripEE-finite-T}) to linear order in $t_0$, taking into account the shift in $z_h$
to ensure that the perturbed solution keeps the original temperature.
This yields
\begin{align}\label{eqn:EE-strip-large-old-T}
 \mathcal S^{(1)}_\mathrm{EE}&=2\pi T_0 V_{d-1}(d-1)z_h^{1-d}\left(j(z_h)-h(z_h)+c_2\right)~.
\end{align}
We compare this to the thermodynamic entropy $\mathcal S=V_{d-1}^{-1}(-\partial\Omega/\partial T)_{\mu,V_{d-1}}$.
The thermodynamic potential $\Omega$ and chemical potential $\mu$ were given in \cite{Karch:2008fa}
and the results, as illustrated in Fig.~\ref{fig:entropy}, agree nicely.

\section{Discussion}\label{sec:discussion}
In this work we have studied entanglement among the charge carriers in the quantum liquid phase of
flavored \N{4} SYM identified in \cite{Karch:2008fa}.
A specific question in this context is whether there is a logarithmic enhancement
of the area law signaling the presence of a Fermi surface.
Given the non-vanishing ground state entropy and the somewhat peculiar scaling
of the specific heat, the entanglement properties also add an interesting piece of information
to characterize the phase, even where they are as expected.
More generally speaking, our investigation provides an example where results on the
entanglement entropy for a non-trivial state can be obtained in closed form.
For the sphere and the strip we have given it
in (\ref{eqn:deltaS-double-integral-2}) and (\ref{eqn:Delta-strip-2}), respectively.
For small regions, these follow the universal behavior predicted in \cite{Bhattacharya:2012mi,Wong:2013gua}.
The more interesting feature are certainly the volume terms dominating the renormalized entanglement entropy for large regions.
Their presence by itself is not very surprising in view of the non-vanishing entropy density
of the global state\footnote{The state is different in that respect from those studied in \cite{Alishahiha:2012ad,Kulaxizi:2012gy}.
This also violates one of the assumptions in \cite{Swingle:2011np}, and may explain why their argument for
an at most logarithmic enhancement of the area law does not apply.
Finally, we note that the finite entropy does not seem to signal an instability \cite{Ammon:2011hz}.}.
In fact, extensive terms were anticipated in a similar context already in \cite{Barbon:2008sr,Swingle:2009wc},
where they were due to the fact that the bulk theory was capped off in the IR.
However, both of these considerations would lead to the expectation of universal volume terms with
a trivial shape dependence.
What we have found, on the other hand, is that in addition to these expected terms there are contributions with
non-trivial shape dependence.
As the entropy of the translationally invariant global state can hardly account for the shape dependence,
we conclude that these are actually a feature due to entanglement correlations.
While extensive entanglement entropy seems to be a generic feature of random states
in many-body systems \cite{Page:1993df},
this is crucially different in local quantum field theories.
The presence of only short-range interactions typically reduces the amount of entanglement to an area law
for the entanglement entropy in the latter.
The clear-cut implementation of the area law was in fact one of the immediate validations for the holographic
prescription of \cite{Ryu:2006bv}.
Known exceptions are systems with a Fermi surface, where a logarithmically enhanced area law of the form
$(k_F\ell)^{d-2}\log k_F \ell$ would be expected \cite{2010PhRvL.105e0502S,Ogawa:2011bz,Huijse:2011ef}, and non-local
theories \cite{Barbon:2008ut, Fischler:2013gsa,Karczmarek:2013xxa,Shiba:2013jja,Pang:2014tpa}.
In spin-chain models the presence of extensive terms for certain excited states could indeed be linked to the \mbox{(non-)locality}
properties of the Hamiltonian for which the particular state would be a ground state \cite{2009JSMTE..10..020A}.
Our setup, on the other hand, does not seem to straightforwardly fit into either of these categories.
This certainly makes the high degree of entanglement among the charge carriers signaled by the extensive terms
a remarkable feature.
It immediately raises the question for the precise nature of the correlations, 
and may eventually help explain the departure from Fermi liquid behavior.

For finite temperature we have shown that the non-trivial shape dependence of the volume terms only persists
up to a scale set by the inverse temperature.
Beyond that scale the renormalized entanglement entropy settles on the shape-independent volume law contribution
due to the entropy of the global state, as shown in Fig.~\ref{fig:finiteT}.
This seems nicely consistent with the interpretation that the shape-dependence of the volume terms is indeed due
to long-range entanglement correlations, which on large enough scales are simply washed out by thermal fluctuations.
We close with a more detailed discussion of Fermi surfaces. The indicator would be a logarithmically
enhanced area contribution, and we would expect this feature for a Fermi liquid.
Since we have considered a state of non-zero entropy density, one could have expected the
enhancement of the area law to just be hidden behind the shape-independent extensive terms.
The expansion (\ref{eqn:SEE-sphere-large-q}), however, shows that the first subleading term
already is a pure area term with no logarithmic enhancement.
The area contribution is thus just enhanced by a shift in the coefficient.
On the other hand, as argued above, the volume terms are non-trivial and an
entanglement effect, so there actually is a more drastic enhancement instead.
As discussed in Sec.~\ref{sec:D3/D7-gauge-field},
the linearized approximation becomes unreliable in the deep IR limit.
A conservative interpretation of our zero-temperature analysis therefore is as a
limit where $T$ is positive but small enough to be neglected on the scales on which
we have studied entanglement correlations.
This picture is confirmed by our finite-temperature analysis,
showing that the volume terms are not just an artefact.
Without the fully backreacted geometry we can, however, not draw definite conclusions on
the limit $q\ell^{d-1}\rightarrow\infty$ and the presence of a Fermi surface.

\begin{acknowledgments}
  We thank Kristan Jensen, Matthias Kaminski and Andy O'Bannon for helpful discussions. 
  HC and CFU thank the organizers of ``Strings 2014'' for the nice conference and the 
  participants for many interesting discussions.
  The work of HC and AK is supported in part by the US Department of Energy under grant number DE-SC0011637.
  CFU is supported by {\it Deutsche Forschungsgemeinschaft} through a research fellowship.
\end{acknowledgments}

\bibliography{flavorbranes.bib}
\end{document}